# Typicality in Ensembles of Quantum States: Monte Carlo Sampling vs Analytical Approximations

**Barbara Fresch, Giorgio J. Moro**

*Department of Chemical Science, University of Padova, Via Marzolo 1, 35131 Padova – Italy*

## Abstract

Random Quantum States are presently of interest in the fields of quantum information theory and quantum chaos. Moreover, a detailed study of their properties can shed light on some foundational issues of the quantum statistical mechanics such as the emergence of well defined thermal properties from the pure quantum mechanical description of large many body systems. When dealing with an ensemble of pure quantum states, two questions naturally arise: what is the probability density function on the parameters which specify the state of the system in a given ensemble? And, does there exist a "most typical" value of a function of interest in the considered ensemble?

Here two different ensembles are considered: the Random Pure State Ensemble (RPSE) and the Fixed Expectation Energy Ensemble (FEEE). By means of a suitable parameterization of the wave function in terms of populations and phases, we focus on the probability distribution of the populations in these ensembles. A comparison is made between the distribution induced by the inherent geometry of the Hilbert Space and an approximate distribution derived by means of the minimization of the informational functional. While the latter can be analytically handled, the exact geometrical distribution is sampled by a Metropolis-Hastings algorithm. The analysis is made for an ensemble of wavefunctions describing an ideal system composed of $n$ spins $1/2$ and reveals the salient differences between the geometrical and the approximate distributions. The analytical approximations are proven to be useful tools in order to obtain ensemble averaged quantity. In particular we focus on the distribution of the Shannon entropy by providing an explanation of the emergence of a typical value of this quantity in the ensembles.

## 1. Introduction

The characterization of thermodynamic properties as well as the study of the dynamics of many body quantum systems are central issues of physical chemistry which should be treated within a suitable statistical framework. The need of a statistical approach to these long standing problems can be understood from various points of view. On a methodological ground there are no doubts that statistical methods are the right conceptual tools to build the bridge between a pure (quantum) mechanical description of complex systems and their characterization in terms of thermodynamic quantities or relaxation behaviour. Still, from a more practical standpoint, one has to face the difficulties in performing efficient numerical simulations of large quantum systems due to the well know exponential scaling of the required computation resources with the system dimension[1]. Moreover, although several schemes have been developed in order to increase the quantum simulation performances[2,3,4] one should also consider a more subtle difficulty which finds its roots in some foundational aspects of the standard quantum statistical mechanics[5,6]. In particular, the following question arises: how statistical mechanics *emerges* from the underlying quantum mechanical description? There is no obvious answer to this question. Even if we could perform a simulation of a large many body quantum system in analogy to a classical molecular dynamics experiment, there is not a straightforward relation between the results of such a calculation, i.e. the time dependent wavefunction of the isolated system, and the standard quantum statistical description based on the statistical density matrix. Of course, one could identify the latter quantity with the time average of the instantaneous density matrix determined by the wavefunction, but in such a case it is not clear why an evolving isolated system should necessarily leads to the microcanonical statistical density matrix[7].

Recently, the possibility of investigating single molecule, or single spin observables[8,9], as well as the necessity of a better understanding of the mechanisms underlying quantum dynamics in order to obtain nanoscale devices and nanostructured materials suitable for quantum computing tasks[10,11,12], have revived the interest in the foundations of quantum statistical mechanics. Important contributions in this field[13,14,15] invite to look at quantum statistical mechanics from a different standpoint. One of the key ingredients of this new perspective consists of shifting the focus from the ensemble averages of the traditional quantum statistics back to the role and predictability of one single realization of a system and its environment described by a quantum mechanical pure state associated to a wavefunction.

Within this perspective the concept of typicality[16] as the key to the emergence of standard statistical equilibrium behaviour has recently been discussed in various works[13,14,15]. In its widest meaning the term typicality indicates that by selecting a set of states on the basis of some conventional rules one obtains a very narrow distribution of some relevant features, which become, actually, typical among those states. In particular it has been shown that a quantum subsystem which is part of a much bigger system described by a pure state can be very likely described by a typical reduced density matrix, which is the canonical one under certain conditions[14]. Nonetheless the emergence of typical values for a large class of observables is not restricted to a small subsystem as it is discussed by Reimann in Ref.15.

The main goal of the present contribution is the development of methodologies for the study of statistical ensembles of quantum pure states. These methodologies are useful tools for the analysis of typicality not only of asymptotically large systems, but also for finite systems amenable to exact numerical computations. They include the following items:

i) The derivation of the probability density (distribution) for ensembles of quantum pure states on the basis of the geometrical measure in the Hilbert space given a suitable parameterization of the wavefunction.

ii) For a given probability density, the generation of a numerical sampling of the ensemble, which allows one to calculate the distribution of any property of the quantum system and, in particular, to assess its typicality.

iii) The derivation of approximate probability distributions leading to a direct calculation of the averages within the ensemble without the need of the numerical sampling.

We emphasize the importance of the third item for the study of large quantum systems, whose characterization through numerical sampling is not feasible. Obviously, the evaluation of typicality for asymptotically large systems falls in such a category of problems. On the other hand, a stronger confidence on the applicability of the invoked approximations could derive from the comparison with the exact numerical results for finite systems.

In the present paper these new methodological tools will be applied to two different ensembles of quantum pure states. In the first one, we consider quantum pure states chosen at random with respect to the uniform measure on the unit sphere in a finite dimensional Hilbert space[17]. They are a much studied subject in the field of quantum information science since they are a resource for a variety of quantum information protocols[18,19]. For example, their entanglement properties have been studied by many authors[20,21,22,23] and several schemes has been proposed for generating them efficiently[24]. Furthermore, ensembles of random states have been previously considered in several studies of the foundations of statistical mechanics[14,25,26]. A statistical sample of states drawn according to such a distribution will be called the Random Pure State Ensemble (RPSE).

In the second type of ensembles, the wavefunctions are chosen at random but with the constraint of having a common expectation value of some observable. We specifically consider the case of fixed expectation value of the energy, but the methodology used by us to obtain the corresponding probability density is rather general and it can be employed to generate any ensemble of this kind[27]. The statistical sample for such a distribution will be denoted as the Fixed Expectation Energy Ensemble (FEEE). This type of distribution has been recently proposed as the quantum counterpart of the classical microcanonical ensemble[28,29,30].

In order to test the capability of these methods to quantify the typicality of a given observable, we shall explicitly consider the Shannon entropy as an important collective property of the quantum system. We emphasize, however, that the same methods can be applied to any chosen property of the quantum system.

In Section 2, we introduce a suitable parameterization of the wave function in terms of populations and phases and derive the corresponding probability density from the inherent geometry of the Hilbert space. In Section 3, the distributions for the considered ensembles are investigated by employing Monte Carlo sampling techniques. Albeit we consider systems of non interacting spins for the sake of simplicity, the procedure can be easily applied to arbitrary quantum systems. In Section 4 we shall derive and discuss approximations of probability distributions for the ensembles, which can be analytically handled, and their validity can be assessed by comparison with the Monte Carlo results.

## 2. Theory

### 2.1 Definition of the Ensembles

Before providing the formal definition of the examined Ensembles of pure states, it is worth to spend some words to clarify how the concept of ensemble is intended in the following. By ensemble we always mean an abstract construction for the statistical sampling of the possible pure states (each of them described by a wavefunction) of an isolated quantum system. Once the states (wavefunctions) have been properly parameterized, one can introduce the probability density with respect to these parameters, and the ensemble becomes simply a realization for the statistical sampling of the probability distribution.

A given isolated quantum system is characterized by its time independent Hamiltonian $H$ which defines the corresponding eigenenergy (orthonormal) basis $\{e_k, k=1,...N\}$, where $N$ denotes the dimension of the Hilbert space $\mathcal{H}$ of the system. Here we shall consider a Hilbert space of arbitrary large but finite dimension. Thus, any wavefunction $\psi$ in such a space can be specified as the superposition of the eigenenergy vectors:

$$\psi = \sum_{k=1}^{N} c^k e_k \tag{1}$$

The $N$-dimensional complex vector $c \equiv (c^1, c^2, ..., c^N) \in \mathbb{C}^N$ can be written as a $2N$ dimensional real vector $x \equiv (\operatorname{Re} c, \operatorname{Im} c) \in \mathbb{R}^{2N}$, and in the latter representation the norm of the state vector is given as Euclidean norm

$$\|\psi\| = \sqrt{\langle \psi | \psi \rangle} = \sqrt{\sum_{j}^{2N} (x^j)^2} \tag{2}$$

Thus, the $\mathbb{R}^{2N}$ space "contains" all the possible non normalized state vectors.

We define the Random Pure State Ensemble (RPSE) as the set of normalized wavefunctions which are uniformly distributed in the Hilbert space $\mathcal{H}$ of the system. The normalization condition

$$\langle \psi | \psi \rangle = 1 \tag{3}$$

defines, according to eq. (2), a $(2N-1)$-dimensional sphere embedded in the $(2N)$-dimensional $\mathbb{R}^{2N}$ space

$$S^{2N-1} = \left\{ x \in \mathbb{R}^{2N} \left| \sum_{j=1}^{2N} (x^j)^2 = 1 \right. \right\} \tag{4}$$

For the set of pure states there is a unique measure which is invariant under all unitary transformations[17], in correspondence to the uniform distribution over the $S^{2N-1}$ unit sphere defined above.

Recently a generalized microcanonical ensemble has been proposed[28,29], with the microcanonical energy identified with the expectation value of the Hamiltonian

$$E = \langle \psi | H | \psi \rangle = \sum_{k=1}^{N} E_k |c^k|^2 \tag{5}$$

$E_k$ being the $k$-th Hamiltonian eigenvalue: $He_k = E_k e_k$. We define the Fixed Expectation Energy Ensemble (FEEE) in correspondence of the set of all the normalized wavefunctions in $\mathcal{H}$, which are characterized by a given value $E$ of the expectation energy according to eq. (5).

For the geometrical analysis of the previously defined Ensembles, it is convenient to introduce the following parameterization of the wavefunctions

$$\psi = \sum_{k=1}^{N} \sqrt{P_k} \exp[i\alpha_k] e_k \tag{6}$$

where the complex coefficients of eq. (1) are written in their polar form. We shall refer to the absolute squares of the coefficients as "populations", $P_k \equiv |c_k|^2$, while the "phases" $\alpha_k$ are defined as the inverse tangent of the ratio between the imaginary and the real part of the coefficients: $\alpha_k \equiv \tan^{-1}(\operatorname{Im} c_k / \operatorname{Re} c_k)$. It should be emphasized that in the space $\mathbb{R}^{2N}$, the set of populations is not normalized, $\sum_k P_k \neq 1$, unless $\|\psi\| = 1$.

## 2.2 Probability Distribution on the Ensembles

In this section we shall sketch the derivation of the ensemble probability distributions on a chosen coordinate set which specifies the wavefunction. The natural underlying assumption is that the probability of a certain set of wavefunctions is proportional to the measure of the set of their representative points in the space where the ensemble is defined.

Let us start by recalling that in the $\mathbb{R}^{2N}$ space the state vector has an Euclidean norm, eq. (2), when the set of coordinates $x \equiv (\operatorname{Re} c, \operatorname{Im} c)$ is used. Therefore, for such a representation an Euclidean geometry can be assumed with an unit metric tensor[31]

$$g_{ij}(x) = \delta_{ij} \tag{7}$$

and the measure of any region is obtained by integration of the elementary volume element $dV = dx = dx^1 dx^2 ... dx^{2N}$. The invariance of the volume measure allows one to write the volume element in arbitrary coordinates $y = (y^1, y^2, \cdots, y^{2N})$ as

$$dV = \sqrt{|g(y)|} dy \tag{8}$$

where $dy = dy^1 dy^2 ... dy^{2N}$, and $g(y)$ is the determinant of the metric tensor $g_{ij}(y)$ in the $y$ representation

$$g_{ij}(y) = \sum_{i',j'=1}^{2N} \frac{\partial x^{i'}}{\partial y^i} g_{i'j'}(x) \frac{\partial x^{j'}}{\partial y^j} = \sum_{j'=1}^{2N} \frac{\partial x^{i'}}{\partial y^i} \frac{\partial x^{i'}}{\partial y^j} \tag{9}$$

The explicit use of the metric tensor is convenient because it permits to determine the proper volume element of any surface embedded in the original Euclidean space[31]. In general, $n$ conditions on the coordinates $y$ of a $2N$ dimensional space define an hyper surface of dimension $K = 2N - n$. In the neighbourhoods of any non singular point of the surface, one can introduce a local set of coordinates $z = (z^1, ..., z^K)$ which parametrically determines the location of a point on the surface: $y^i(z)$ for $i = 1, 2, \cdots, 2N$. The metric tensor induced on the surface is then given by

$$g_{kk'}(z) = \sum_{ij=1}^{2N} \frac{\partial y^i}{\partial z^k} g_{ij}(y) \frac{\partial y^j}{\partial z^{k'}} \tag{10}$$

where $k, k' = 1, ..., K$. Once the metrics on the surface is known, like in eq. (8) one can specify the corresponding volume element as $dV = \sqrt{|g(z)|} dz$ with $dz = dz^1 dz^2 ... dz^K$.

Since both ensembles, RPSE and the FEEE, are defined by constraints which define hyper surfaces embedded in the $\mathbb{R}^{2N}$ space, one can apply the above mentioned scheme to obtain the corresponding volume element $dV$. Then, because of the assumed proportionality between the probability that the state vector belongs to a given set and its geometrical measure, one can specify the probability density $p(z)$ with respect to coordinates $z$ according to the relation

$$p(z)dz = \frac{dV}{\int_D dV} \tag{11}$$

where the normalization is calculated as the total volume of the set $D$ for the allowed pure states. In this way, from the measure, one derives the probability density of the surface points

$$p(z) = \frac{\sqrt{|g(z)|}}{\int_D \sqrt{|g(z)|} dz} \tag{12}$$

The representation of the $\mathbb{R}^{2N}$ space by means of populations and phases, $y = (P, \alpha)$ where $P = (P_1, P_2, \cdots, P_N)$ and $\alpha = (\alpha_1, \alpha_2, \cdots, \alpha_N)$, is particularly convenient since the constraints of state vector normalization eq. (3) and fixed expectation energy eq. (5) involve populations only

$$\sum_k P_k = 1 \tag{13}$$

$$\sum_k P_k E_k = E \tag{14}$$

This implies that the set of populations are statistically independent from the phases, and also that each phase variable is characterized by an uniform distribution in the corresponding definition domain $[0, 2\pi)$ (see Appendix A for the details). Thus, the probability density on the populations $P$ and phases $\alpha$ can be factorized as

$$p(P, \alpha) = p(P) p(\alpha) = p(P) / (2\pi)^N \tag{15}$$

For the Random Pure State Ensemble (RPSE) only the condition (13) has to be considered and thus one can use all the set of phases $\alpha$ and the $(N-1)$ independent populations $P = (P_1, ... P_{N-1})$ as coordinates $z$ on the corresponding surface. Appendix A describes in detail the calculation of the determinant $g(z)$ of the metric tensor leading to the following RPSE probability density on the populations

$$p_{RPSE}(P_1, ..., P_{N-1}) = (N-1)! \tag{16}$$

The set of populations which characterizes wavefunctions in the Fixed Expectation Energy Ensemble (FEEE) has to satisfy both eq. (13) and eq. (14). If the two constraints are used to determine the populations $P_N$ and $P_{N-1}$, then as described in Appendix A one derives the following explicit form for the probability distribution

$$p_{FEEE}(P_1, ..., P_{N-2}) = \frac{1}{C} \left[ \sum_{j=1}^{N-2} P_j (1+a_j) a_j - \left( \frac{E - E_{N-1}}{E_N - E_{N-1}} \right)^2 + \left( \frac{E - E_{N-1}}{E_N - E_{N-1}} \right) \right]^{1/2} \tag{17}$$

with coefficients $a_j = \frac{E_{N-1} - E_j}{E_N - E_{N-1}}$, and the constant $C$ determined by the normalization of the probability density.

By using the ensemble distribution, eq. (15), with the corresponding probability density on populations, eq. (16) or eq.(17), one can obtain in principle the ensemble average of any function of a quantum state $a(|\psi\rangle) = a(P, \alpha)$

$$\langle a \rangle = \int_D a(\alpha, P) \, p(\alpha, P) dP d\alpha \tag{18}$$

as well as the average of its higher moments

$$\langle a^n \rangle = \int_D a(\alpha, P)^n \, p(\alpha, P) dP d\alpha \tag{19}$$

Amongst the possible functions, one can include the expectation value $a = \langle \psi | A | \psi \rangle$ of an operator $A$ defined in the Hilbert space. It should be emphasized that, if these averages are available, then one can quantify the typicality of a property described by a real function $a(P, \alpha)$ by examining its variance

$$\sigma_a \equiv \sqrt{\langle a^2 \rangle - \langle a \rangle^2} \tag{20}$$

In conclusion, for the two types of ensemble here introduced we have derived the corresponding probability distributions which in principle allow the calculation of the average of any property and to characterize the typicality as well. This, however, requires the evaluation of multidimensional integrals that rarely can be performed directly. Therefore, in order to characterize the statistical behaviour of a property within an ensemble, one has to introduce a suitable approximation of the probability distribution that allows an analytical estimate of the averages, or to perform a statistical sampling for the given probability distribution. The first route will be tackled in Section 4, while in the next Section we address the issue of the statistical sampling. Albeit the procedures we present are general enough to analyze the typicality of any property, in order to provide specific examples of their application, we shall consider the following entropy function

$$S = -\sum_{k=1}^{N} P_k \ln P_k \tag{21}$$

It corresponds to the Shannon entropy in the energy representation, which is usually interpreted as a measure of the lack of information about the outcome of the measurement of the energy. In the present framework we do not consider the measurement process, so that the function (21) is rather interpreted as a measure of the degree of disorder of a quantum pure state in relation to its decomposition into the Hamiltonian eigenstates. In particular a vanishing entropy would be recovered only for a stationary state, $\psi \propto e_k$ for a given eigenstate $e_k$. It should be noted that several theoretical entropy measures for quantum systems have been discussed in the literature[32,33,34], nonetheless the issue of the relation between such informational measure and the thermodynamic entropy is still subject of a lively debate and active research[35,36,37,38].

## 3. Monte Carlo Ensemble Sampling and Typicality

We shall employ Monte Carlo sampling techniques in order to draw a statistical sample of population sets and study the corresponding Ensemble distributions of the populations and of the entropy function eq. (21). Ensembles of spins are convenient systems for investigations of quantum statistical behaviour, since one has to consider a finite dimensional Hilbert space. This kind of model systems is the subject of a continuously increasing attention either from a theoretical[11,39] as well as experimental perspective[8,9,40] because it represents the natural test bed for quantum information protocols. We thus consider a system composed of $n$ non interacting $1/2$ spins, each spin having its Zeeman frequency $\omega_k$, so that the total number of states of the system is $N = 2^n$.

### 3.1 Population and Entropy Distribution in the RPSE

In order to study the Random Pure State Ensemble (RPSE) one has to draw samples from the uniform probability distribution on the $N-1$ simplex, eq. (16), of the populations. The problem can be solved by introducing an auxiliary set of variables $\xi \equiv (\xi_1, ..., \xi_{N-1})$ uniformly distributed in $(0,1]$. It is easily shown[26,41], that the set of populations calculated as

$$P_1 = 1 - \xi_1^{\frac{1}{N-1}}, \quad ......, \quad P_J = \left(1 - \xi_J^{\frac{1}{N-J}}\right) \prod_{i=1}^{J-1} \xi_i^{\frac{1}{N-i}}, \quad ......, \quad P_N = \prod_{i=1}^{N-1} \xi_i^{\frac{1}{N-i}} \quad (22)$$

is a realization from the RPSE distribution, eq. (16). This allows one to efficiently generate a statistical sample of the population set for the RPSE and thus the corresponding distribution of the entropy through the definition (21).

Figure 1 shows the marginal distribution of a single population obtained as (normalized) histogram of the statistical sample numerically generated from the geometrical distribution, eq. (16). It should be evident that RPSE, whose probability distribution is invariant with respect to the exchange of population variables, does not privilege any population and, therefore, leads to the same marginal distribution for all the populations. It is also clear that an increase of the number of spins moves the marginal distribution to lower values of the population variable. This is a direct consequence of the dependence of the average population $\langle P_k \rangle = 1/N$ on the dimension $N$ of the Hilbert space, simply deriving from the population normalization. A less obvious behaviour which appears evident from the diagrams of Figure 1, is that the population distribution has a width $\sigma_{P_k}$ always comparable to the population average $\langle P_k \rangle$. Such a feature, which will be rationalized in Section 4, brings to the conclusion that within RPSE the populations do not display typicality.

On the other hand from the standard quantum statistical mechanics one has the intuition that at least some functions of the state of the system, such as the entropy, should not depend on the details which specify the state, i.e. on a particular choice of the population set, but only on "gross properties" of the quantum state. In Figure 2, the Ensemble Distribution of the entropy per spin, i.e. $S/n$, for systems composed of different numbers of spins, is reported. The histograms resulting from the sampling are well fitted by Gaussian distributions. In the inset of the figure the standard deviations of the fitting distributions are reported as a function of the number of spins in the system.

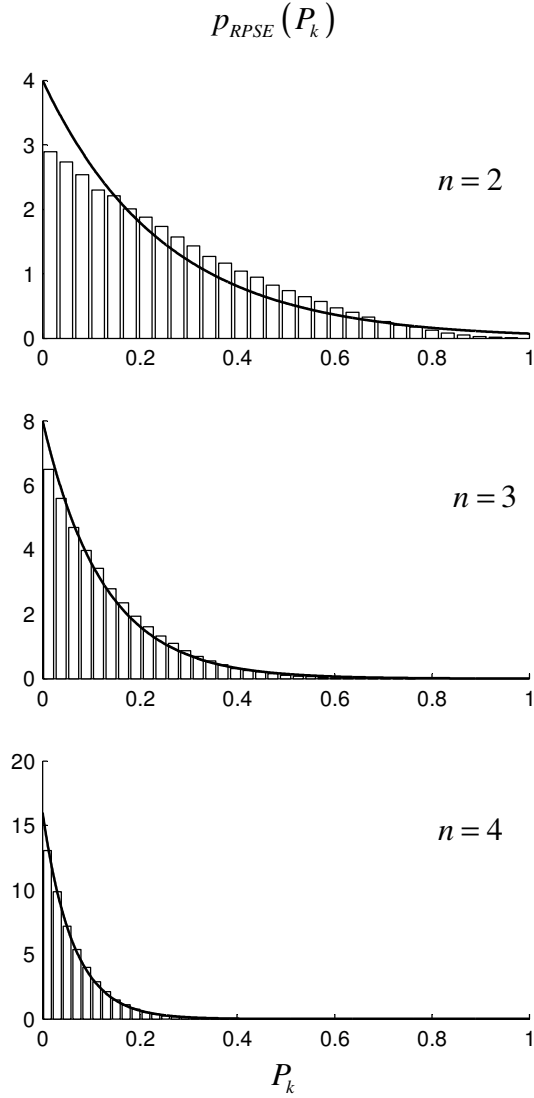

**Figure 1: Marginal distribution of a population in the RPSE: the (normalized) histograms refers to the statistical sample of $10^5$ points (in 30 bins) generated by the algorithm described in the text, for a system composed of $n = 2, 3, 4$ spins respectively. The continuous lines represent the exponential distribution, eq. (29) discussed in Section 4.1.**

The main point which emerges from these calculations is the existence of a typical value of the entropy per spin in the considered ensemble of pure states. Indeed, as the number of spins increases, the entropy distribution becomes a sharper and sharper function peaked around a typical value which can be identified with the average entropy. Such a property which is called typicality has been recently used to explain the emergence of the canonical state for a subsystem which is part of a much bigger system described by a pure state[13,14]. Reimann[15] has proven that under certain hypothesis the property of typicality also holds for the expectation value of a generic observable. Our results on the RPSE distribution of the entropy have to be interpreted as an evidence of the property of typicality. In other words, as the size of the system increases, the vast majority of the different pure states which belong to the RPSE are characterized by nearly the same value of the entropy function even if each of them is defined by a different set of populations.

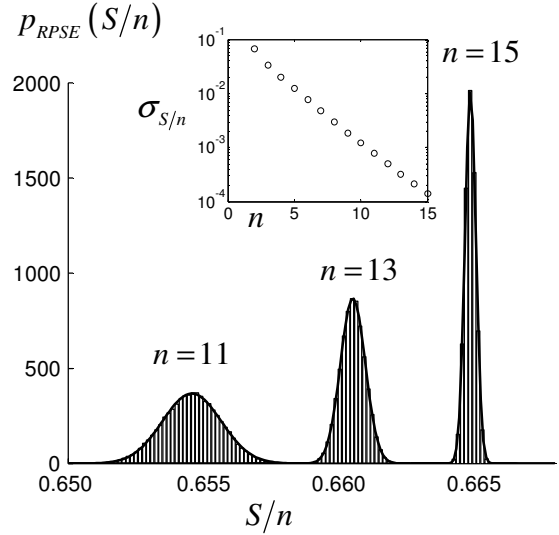

**Figure 2: Distribution of the entropy per spin $S/n$ for different numbers of spins as obtained by numerical sampling ($10^5$ sampled points) of the RPSE distribution of systems composed by $n = 11, 13, 15$ spins as indicated in the figure. In the inset the standard deviation of the fitting Gaussian distributions are reported as a function of the number of spins which constitutes the system.**

### 3.2 Population and Entropy Distribution in the FEEE

For studying the Fixed Expected Energy Ensemble (FEEE) we need to generate a statistical sample from the probability density eq. (17). Since the normalization constant is unknown and we deal with a multivariate probability distribution, the use of a Markov Chain becomes convenient. Our sampling of the FEED has been performed by means of a Metropolis-Hastings algorithm[42,43]. In particular, a random walk updating scheme[44] has been implemented[7].

The algorithm is basically made up by the following steps

1. Start at an arbitrary point $X$ which belong to the domain of the target distribution $p_{FEEE}(P_1,...,P_{N-2})$, with $p_{FEEE}(X) > 0$

2. Generate a random variable $Y$ from an arbitrary but fixed proposal distribution $q(X,Y)$, this represents a proposed move from the state $X$ to the state $Y$. In the random walk updating scheme the proposed new value $Y$ equals the current values $X$ incremented by a random variable $Z$: $Y = X + Z$. In this case $q(X,Y) = g(Y-X) = g(Z)$ is the probability density associated to the random variable $Z$. We have used the multivariate Gaussian distribution

$$g(Z) = \frac{1}{(2\pi)^{N/2} (\det \Sigma)^{1/2}} \exp\left[-\frac{1}{2} Z^T \Sigma^{-1} Z\right] \qquad (23)$$

where $\Sigma$ is the covariance matrix. A diagonal covariance matrix was used, with each entry proportional to the average populations calculated on the basis of the approximate distributions introduced in Section 4

$$\Sigma_{kk'} \propto \delta_{kk'} (\lambda + \mu E_k)^{-1} \qquad (24)$$

where the $\lambda$ and $\mu$ parameters are defined in eq. (36). The variance of the proposal distribution can be thought of as a set of tuning parameters to be adjusted to get an optimal sampling of the target distribution.

3. Calculate what can be termed the probability of move

$$\alpha(X,Y) = \min\left[1, \frac{p_{FEEE}(Y)}{p_{FEEE}(X)}\right] \qquad (25)$$

4. Generate a random variable $u$ uniformly from $[0,1]$: if $u < \alpha$ accept the proposal and move to state $Y$, otherwise reject the proposal and remain in $X$, to be considered as the new sample. Repeat 1-4.

In order to determine the FEEE target distribution, one has to specify the energy spectrum of the considered system. Let $|M\rangle = |m_1^M m_2^M .... m_n^M\rangle$ be an eigen-energy state of the system composed of $n$ non interacting $1/2$ spins. Thus, the corresponding energy is given as $E_M = \sum_{k=1}^{n} \hbar \omega_k m_k^M$, where $m_k^M = \mp 1/2$ and $\omega_k$ is the Zeeman frequency of the $k$-th spin. We have to study the distribution of the populations and of the entropy as a function of the expectation energy $E = \sum_{M=1}^{N} P_M E_M$. In practice it is convenient to employ the energy *per spin* $\varepsilon = E/n$ as the only independent parameter which defines the FEEE of a given system. For the sake of simplicity the distributions here reported refer to a system of $n$ identical spins, i.e. $\omega_k = \omega_0$ for all $k$; however, it has been verified that the general features of resulting distributions do not depend on such a particular assumption. In the following $\hbar \omega_0$ will be used as the energy unit, and we conventionally set the zero of the energy scale in correspondence of the ground state, i.e. $E_1 = 0$. Figures 3 and 4 report the distributions of single populations obtained from the numerical. Due to the presence of the expectation energy constraint, the populations corresponding to different energy levels are not statistically equivalent. In Figure 3 the distributions of the population $P_2$, corresponding to $E_2 = 1$, and of the population $P_{43}$, corresponding to the energy eigenvalue $E_{43} = 4$, are shown. These calculations refer to a system composed of $n = 6$ spins with energy per spin equal to $\varepsilon = 0.2$. Even if the populations are not statistically equivalent, each marginal distribution has an exponential-like profile like in the RPSE. This is generally true for all the populations, and for all the possible values of the energy parameter $\varepsilon$, but with the notable exception of the ground state population. As it is evident from Figure (4), the ground state population has a distribution with a peaked profile, at least for not too large expectation energies. Furthermore, the centre of such a distribution moves toward the unity with a decrease of its width for $\varepsilon \to 0$. Indeed, to attain such a limit, the ground state population should increase up to reach its maximum value.

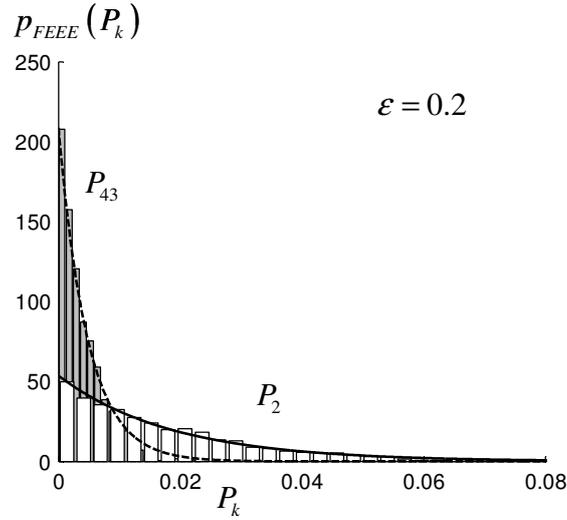

**Figure 3: Marginal distributions of $P_2$, corresponding to the energy eigenvalue $E_2 = 1$, and of $P_{43}$, corresponding to the energy eigenvalue $E_{43} = 4$, obtained from the numerical sampling ($2 \cdot 10^5$ sampled points) of the FEEE for a system of $n = 6$ spins. The expectation energy per spin is set to $\varepsilon = 0.2$. The solid lines represent the approximate distributions discussed in Section 4.2.**

Figure 5 shows the FEEE distribution of the entropy per spin obtained from the Monte Carlo sampling for different values of the expectation energy per spin. The upper panel refers to a system of $n = 6$ spins while the lower panel is the sampling for $n = 10$ spins. These results clearly show that the entropy distribution is concentrated around a typical value. Furthermore, the comparison of the distributions for $n = 6$ and for $n = 10$ cases clearly demonstrates that the enlargement of the system size produces a narrowing of the entropy distribution, like for the RPSE distributions previously examined. The main difference is that for FEEE the distribution of the entropy and its typical value (i.e., the distribution maximum) depends on the expectation energy.

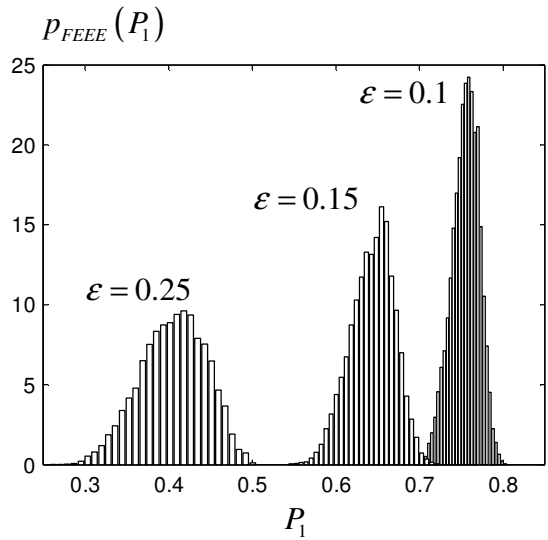

Figure 4: Marginal distributions of the first population $P_1$ corresponding to the ground state. The calculations refer to the FEEE ($2 \cdot 10^5$سampled points) of a system composed by $n = 6$ spins for three different values of the expectation energy per spin as reported in the figure.

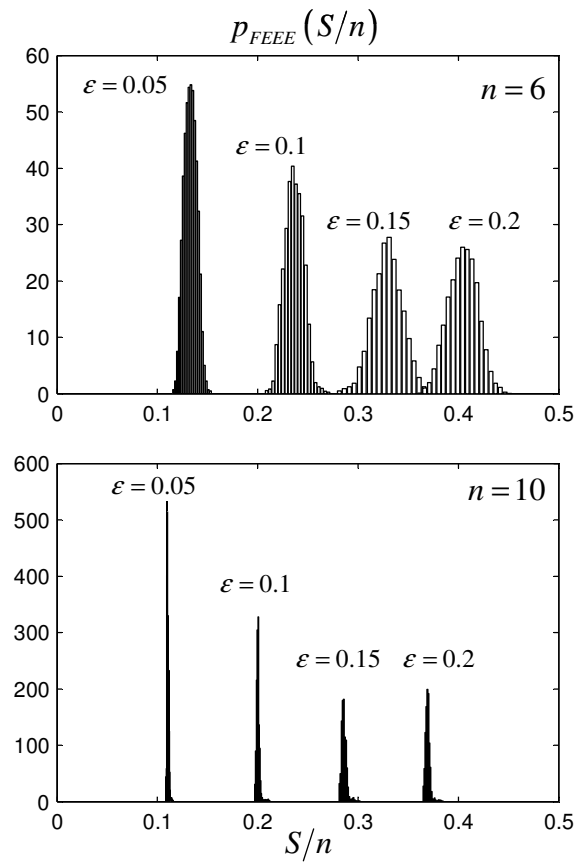

Figure 5: FEEE distributions of the Shannon entropy per spin for a system composed of $n = 6$ spins (upper panel) and $n = 10$ spins (lower panel). The (normalized) histograms refer to different values of the expectation energy per spin as reported in the figure.

## 4 Approximate Distributions from Minimization of the Information Functional

### 4.1 Approximate Distribution for the RPSE

An approximate form of the distribution for the Random Pure State Ensemble has been proposed by Wootters[45]. Such a distribution is very useful since it allows straightforward estimates of the averages in Ensemble, which result to be rather accurate in the comparison with the numerical results of the Monte Carlo calculations. We shall derive the same approximate distribution by following a different method based on the information functional, which can be easily generalized to Ensembles different from the RPSE, like the FEEE, as shown in the following.

The key methodological ingredient of the Wootters (and our) method is the replacement of the populations, which in the RPSE should satisfy the normalization constraint, $\sum_{k=1}^{N} P_k = 1$, with a set of $N$ random variables $\eta = (\eta_1, \eta_2, \cdots, \eta_N)$, each of them defined in the domain $[0, \infty]$, with a distribution characterized through the probability density $W(\eta)$ normalized as

$$\int d\eta\, W(\eta) = \int_0^\infty d\eta_1 \int_0^\infty d\eta_2 \cdots \int_0^\infty d\eta_N\, W(\eta) = 1 \tag{26}$$

and which allows the determination of the average $\langle f(\eta) \rangle$ of any function $f(\eta)$ of the stochastic variables $\eta$. The important point to emphasize is that, in general, variables $\eta$ do not satisfy the normalization condition, $\sum_{k=1}^{N} \eta_k \neq 1$, and, therefore, they cannot be identified with populations. However, the distribution function $W(\eta)$ is chosen in such a way that their average is normalized

$$\sum_{k=1}^{N} \langle \eta_k \rangle = 1 \tag{27}$$

Thus we require that these random variables satisfy *approximately* (that is, on the average) the condition for the RPSE, with *a posteriori* check of the goodness of such an approximation.

The conditions eqs. (26) and (27) do not determine a unique probability distribution function $W(\eta)$. The general problem of the specification of the probability function in the absence of information for its full characterization is as old as the theory of probability itself. The "Principle of Insufficient Reason" of Laplace was an attempt to supply a criterion of choice for the probability if no other stronger reasons are available. The development of information theory and statistical inference has lead to the maximum entropy principle as the rule which permits to determine the least biased distribution according to our initial information[46]. This principle states that, among the infinite set of functions $W(\eta)$ which satisfy the given constraints, it is reasonable to choose the one which minimizes the informational functional

$$I[W] = \int d\eta\, W(\eta) \ln W(\eta) \tag{28}$$

were $I[W]$ denotes a functional dependence of the information content $I$ on the distribution $W(\eta)$. The minimum information (maximum entropy) principle selects the distribution, being compatible with the given constraints, which has the maximum random character in relation to the smoothest dependence of $W(\eta)$ on the stochastic variables $\eta$.

The minimization of the functional (28) under the constraints eqs. (26) and (27) is easily performed by means of the Lagrange Multiplier method. The resulting distribution is factorized into identical exponential distributions for all the components of the random vector $\eta$

$$W_{RPSE}(\eta) = \prod_{k=1}^{N} w_{RPSE}(\eta_k) \qquad w_{RPSE}(\eta_k) = N e^{-N\eta_k} \qquad (29)$$

In conclusion, the random variables (associated to the populations) result to be statistically independent with an exponential distribution for each variable. In Figure 1 one can compare these approximate distributions $w_{RPSE}(\eta_k)$ drawn as continuous lines with the exact marginal distribution of populations deriving from the numerical sampling of RPSE. The convergence of the approximate distribution toward the exact ones clearly emerges for increasing dimensions of the spin system.

On the basis of this result, it is interesting to look at the statistical behaviour of the sum $X = \sum_{k=1}^{N} \eta_k$ of the $N$ random variables $\eta$, whose average is unitary, $\langle X \rangle = 1$, because of the constraint eq. (27). One can apply the Central Limit Theorem[47], to conclude that $X$ is a normally distributed random variable with a variance decreasing with the dimension $N$ of the Hilbert space, as one can explicitly derive from the distribution eq. (29)

$$\left\langle (X - \langle X \rangle)^2 \right\rangle = \langle X^2 \rangle - 1 = \sum_{k=1}^{N} \langle \eta_k^2 \rangle + \sum_{k,k' \neq k}^{N(N-1)} \langle \eta_k \eta_{k'} \rangle - 1 = \frac{1}{N} \qquad (30)$$

In other words, as $N$ become very large, the condition of normalization on the average, eq. (27), becomes *effectively* a condition on the normalization of each realization of the set. However the exact equivalence is found only in the limit $N \to \infty$.

From the statistical independence of the variables in the approximate distribution, eq. (29), one can also justify the Gaussian distribution of the entropy as recovered from the numerical sampling reported in Figure 2. Indeed, if $N$ is large enough, this is the prediction of the Central Limit Theorem when applied to the sum eq. (21) which defines the entropy function. Furthermore, one can directly calculate the average value of the entropy as

$$\langle S \rangle = N \int_{0}^{\infty} d\eta_k w_{RPSE}(\eta_k) \eta_k \ln \eta_k = \ln N - (1 - \gamma) \qquad (31)$$

where $\gamma \cong 0.5772$ is the Euler constant. In order to provide a direct evidence of the accuracy of the approximate result eq. (31) also for small size systems, in Figure 6 we have compared it with the exact numerical values of the average entropy as a function of the number $n$ of spins.

Furthermore, by explicitly calculating the second moment of the entropy distribution one can shown that, at the leading order in $N$,

$$\frac{\sigma_S}{\langle S \rangle} \simeq \frac{1}{\sqrt{N}} \qquad (32)$$

This is equivalent to proving the typicality of the entropy in the RPSE, as long as by increasing the dimension of the Hilbert space, the width of the entropy distribution within the ensemble, when compared to its average, tends to vanish.

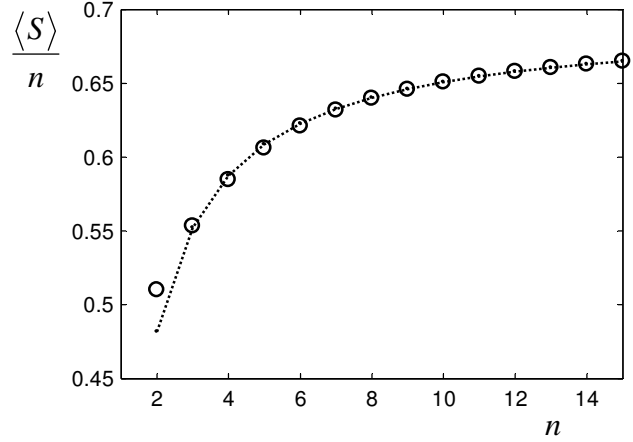

**Figure 6: Average entropy per spin as a function of the number of spins. Circles: numerical sampling of RPSE, dotted line: approximation eq. (31).**

### 4.2 Approximate distributions for the FEEE.

In order to obtain the approximate distribution for the FEEE, the minimization of the informational functional eq. (28) has to be performed under the additional constraint on the average of the energy expectation value

$$\sum_{k=1}^{N} \langle \eta_k \rangle E_k = E \qquad (33)$$

Like for the constraint on the population normalization, within such an approximation the constraint on the expectation energy is taken into account in the mean, by replacing the populations $P_k$ in eq. (14) with the averages $\langle \eta_k \rangle$ of the stochastic variables. In the following we shall assume that the energy eigenvalues are ordered in magnitude, $E_k \leq E_{k+1}$, and that the origin of the energy scale is chosen in correspondence of the lowest energy eigenvalue: $E_1 = 0$. The result, by applying the Lagrange Multipliers method, is again a distribution factorized in $N$ independent exponential distributions

$$W_{FEEE}(\eta) = \prod_{k=1}^{N} w_{FEEE}(\eta_k) \qquad w_{FEEE}(\eta_k) = (\lambda + \mu E_k) e^{-(\lambda + \mu E_k)\eta_k} \qquad (34)$$

but now the averages of the random variables depend on the corresponding energy eigenvalues

$$\langle \eta_k \rangle = (\lambda + \mu E_k)^{-1} \tag{35}$$

The Lagrange parameters $\lambda$ and $\mu$ are implicitly determined by the following equations

$$\sum_{k=1}^{N} \frac{1}{(\lambda + \mu E_k)} = 1 \qquad \sum_{k=1}^{N} \frac{E_k}{(\lambda + \mu E_k)} = E \tag{36}$$

Notice that when the energy expectation value $E$ approaches the value corresponding to all the energy levels equally populated, $E^* = \sum_{k=1}^{N} E_k / N$, then these parameters tend to the limiting values $\lambda \to N$ and $\mu \to 0$. Correspondingly the population distribution $W_{FEEE}(\eta)$ tends to the RPSE population distribution eq. (29).

The distribution eq. (29) for the RPSE and eq. (34) for the FEEE are exactly the distributions proposed by Wootters[45] on the basis of different type of considerations. An important feature of these distributions is their factorization into single variable distributions whose domain is the entire positive real axis. Thus, the ensemble average of any function of the quantum state can be analytically calculated. For example, according to the approximate distribution eq. (34), the FEEE average entropy reads

$$\langle S \rangle = -\sum_{k=1}^{N} \int_0^\infty d\eta_k w_{FEEE}(\eta_k) \eta_k \ln \eta_k = \sum_{k=1}^{N} \frac{\ln(\lambda + \mu E_k)}{\lambda + \mu E_k} - (1-\gamma) \tag{37}$$

which should be considered as an implicit function of the expectation energy (or its value per spin $\varepsilon = E/n$) because of the constraints eqs. (36) for the parameters $\lambda$ and $\mu$.

However, the comparison between the average entropy calculated according to eq. (37) and the average value obtained by the numerical sampling of the FEEE distribution of the entropy for a 10 spin system, points out a discrepancy. This is clearly shown in Figure 7 where the averages of the entropy per spin obtained from the Monte Carlo sampling of the FEEE distribution for different values of the expectation energy per spin $\varepsilon = E/n$ are represented by circles, while the average entropy calculated according to eq. (37) is represented by the dotted line. The difference between the results of the two procedures increases as the energy decreases. In particular eq. (37) predicts a negative entropy for low values of the energy per spin $\varepsilon$, a value which can never be recovered from definition eq. (21) of the entropy whatever is the set of populations.

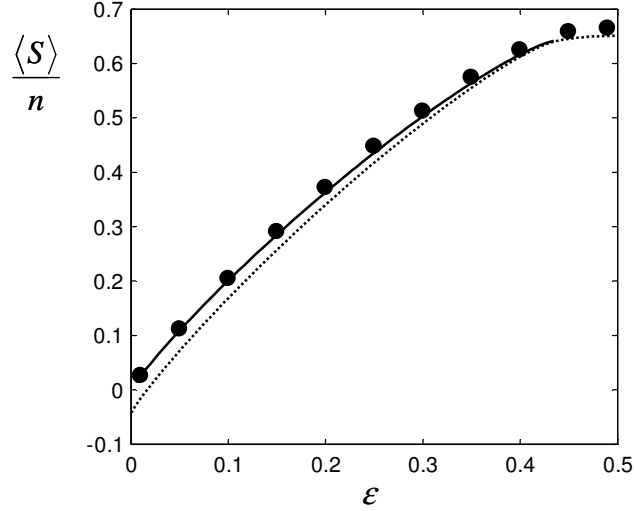

**Figure 7: FEEE average entropy per spin as a function of the expectation energy per spin for a system composed of $n=10$ spins. The figure displays: the average values obtained from Monte Carlo sampling (circles), the average according to the Wootters approximation of the distribution eq. (34), (dotted line), and the average according to the second approximate form of the FEED eq. (42), (continuous line).**

In order to understand the origin of these discrepancies, one should compare the exact population distributions as obtained from the Monte Carlo procedure, with the approximate ones previously derived. We have verified that the exponential distribution given in eq. (34) is a good approximation for all the populations, except for the ground state population $P_1$. As an example, Figure 3 shows the good agreement recovered for the distributions on the populations $P_2$ and $P_{43}$ in a particular situation. On the contrary, no agreement is found for the distribution on $P_1$ because, as shown in Figure 4, in this case the Monte Carlo procedure leads to a distribution with a maximum, which obviously cannot be recovered from the exponential form eq. (34).

In order to obtain a better approximation, we should take into account the peculiarity of the distribution on the ground state population. The basic idea is to use again the procedure of minimization of the information functional eq.(28), but with a specific restriction on the space of the allowed function for the distribution on the first variable $\eta_1$. Then we introduce a new probability distribution, denoted as $W_{FEEE,II}(\eta)$, where the functional dependence on $\hat{\eta} = (\eta_2, \cdots, \eta_N)$ has to be determined from the minimization of the information functional eq. (28), while for the first variable $\eta_1$ (associated to the ground state population $P_1$) we impose a well defined functional form $G_a(\eta_1)$ even if parametrically dependent on a set of constants $a = (a_1, a_2, \cdots)$. Thus the overall probability density is written as

$$W_{FEEE,II}(\eta) = G_a(\eta_1)\hat{W}_{FEEE,II}(\hat{\eta}) \tag{38}$$

where $\hat{W}_{FEEE,II}(\hat{\eta})$ describes the functional dependence on $\hat{\eta}$. For the functional dependence on the first variable, because of the bell shaped profiles suggested by the results of Monte Carlo calculations and reported in Figure 4, we choose a normalized Gaussian function

$$G_a(\eta_1) = \sqrt{\frac{a_2}{\pi}} e^{-(\eta_1 - a_1)^2 a_2} \tag{39}$$

specified according to the two parameters $a = (a_1, a_2)$ for the centre and the inverse squared width of the distribution. As a matter of fact, truly Gaussian distributions are recovered from the Monte Carlo calculations when sufficiently low energy per spin is considered, as one can perceive from Figure 4.

The procedure employed to determine the distribution eq. (38) which minimizes the information functional eq. (28) is described in detail in Appendix B, and here we report only the final result. For the functional dependence on the variables $\hat{\eta}$ an exponential form is recovered like in the previous approximation eq. (34)

$$\hat{W}_{FEEE,II}(\hat{\eta}) = \frac{N-1}{E} \prod_{k=2}^{N} E_k \exp\{-(N-1)E_k \eta_k / E\} \tag{40}$$

while for the Gaussian distribution eq. (39) the following relation is derived for its centre

$$a_1 = 1 - \frac{E}{N-1} \sum_{k=2}^{N} \frac{1}{E_k} \tag{41}$$

and a vanishing value for its width (i.e., $1/a_2 = 0$) which corresponds to a Dirac delta profile. In conclusion the resulting probability density is specified as

$$W_{FEEE,II}(\eta) = \delta(\eta_1 - a_1)\hat{W}_{FEEE,II}(\hat{\eta}) \tag{42}$$

This second form of the approximate distribution is practically equivalent to the previous one for all the populations except the first one, which is distributed like a Dirac delta at its average value $\langle \eta_1 \rangle = a_1$. Such a distribution can be employed to evaluate the average entropy

$$\langle S \rangle = -\sum_{k=1}^{N} \langle \eta_k \rangle \ln \langle \eta_k \rangle - (1-\gamma)(1-\langle \eta_1 \rangle) \tag{43}$$

It can be rewritten as an explicit function of the scaled expectation energy $e = E/N$

$$\langle S \rangle = -(1-eS_0)\ln(1-eS_0) - eS_0 \ln e + eF_0 - (1-\gamma)eS_0 \tag{44}$$

where the quantities $S_0 = \sum_{k=2}^{N} 1/E_k$ and $F_0 = \sum_{k=2}^{N} (\ln E_k)/E_k$ are characteristic properties of the energy spectrum of the considered system. In Figure 6 such an average entropy per spin is represented as a continuous line, and a good agreement is found in the comparison with the numerical sampling of the exact distribution. In particular, the unphysical negative values of entropy are avoided.

However, Monte Carlo results point out that the distribution on $P_1$ has a finite width. Of course this represents a shortcoming of such an approximation which could be overcome by introducing more complex profiles for $G_a(\eta_1)$, but at the price of a much more cumbersome procedure. On the other hand, we think that there is not strictly a necessity of these further developments, as long as the approximate distribution eq. (42) leads to accurate enough values for the average entropy.

## 5. CONCLUSIONS

In order to identify the methodologies suitable to statistically characterize quantum pure states, we have presented an analysis of the probability distributions which emerge from the geometry of the Hilbert space. In particular by using a parameterization of the wavefunction in terms of populations and phases we have explicitly derived the distributions of these variables associated to the Random Pure State Ensemble and to the Fixed Expectation Energy Ensemble. While the phase variables are statistically independent and uniformly distributed, the ensemble probability densities on the populations are defined in high dimensional domains with a non trivial topology, because the populations are not statistically independent for the presence of the constraints. On the one hand we have characterized such distributions through numerical sampling by using Monte Carlo techniques. On the other hand analytical approximations valid in the large $N$ limit have been developed and compared with the numerical results. These methods allow one to study the marginal distributions of single populations, as well as those of any function of interest. We have focused on the Shannon entropy associated to the pure state of the whole system in the energy representation. An interesting point which emerges from the study of the ensemble distribution in the chosen model system composed of $n$ spins $1/2$, is the following: while the probability distribution of the populations itself is the broadest one compatible with the constraints of the considered ensemble, the ensemble distributions of the observable entropy are, on the contrary, very peaked functions, and this allows its characterization trough their typical values. Within the statistical framework presented here, the typicality of the entropy is an evidence which does not depend on the nature of the system, rather it is a simple consequence of the high dimensionality of the phase space together with the structure of the observed function which is defined as a sum of many terms. For this reason the tools developed here can be used to calculate and analyze the typical values of a wide class of observables, including the expectation value of a generic operator or the elements of the reduced density matrix of a subsystem. The connection between the behaviour of such typical values and the relevant constraints which define the ensembles are an interesting issue which will be considered elsewhere.

# APPENDIX A: Metric tensors and volume elements in the Ensemble representative spaces

Let us first consider the space $\mathbb{R}^{2N}$ and perform the change of representation from the Euclidean coordinates $x \equiv (\operatorname{Re} c, \operatorname{Im} c)$ to the generalized coordinates $y = (P, \alpha)$ determined by populations and phases. The Jacobian matrix of the following transformation

$$x_{2i-1} = \sqrt{P_i} \cos \alpha_i \qquad x_{2i} = \sqrt{P_i} \sin \alpha_i \tag{45}$$

for $i = 1, 2, \cdots, N$, is block diagonal and the i-th block reads

$$\begin{pmatrix} \dfrac{\partial x_{2i-1}}{\partial P_i} & \dfrac{\partial x_{2i-1}}{\partial \alpha_i} \\ \dfrac{\partial x_{2i}}{\partial P_i} & \dfrac{\partial x_{2i}}{\partial \alpha_i} \end{pmatrix} = \begin{pmatrix} \left(2\sqrt{P_i}\right)^{-1} \cos \alpha_i & -\sqrt{P_i} \sin \alpha_i \\ \left(2\sqrt{P_i}\right)^{-1} \sin \alpha_i & \sqrt{P_i} \cos \alpha_i \end{pmatrix} \tag{46}$$

Therefore, according to eq. (9), the metric tensor in the $y = (P, \alpha)$ representation is diagonal with components

$$g_{P_i P_i} = \frac{1}{4P_i} \qquad g_{\alpha_i \alpha_i} = P_i \tag{47}$$

from which one derives the volume element in the new set of coordinates

$$dV = \sqrt{|g(P,\alpha)|} \, dP \, d\alpha = 2^{-N} dP_1 \ldots dP_N d\alpha_1 \ldots d\alpha_N \tag{48}$$

By imposing the constraint eq. (13), which can be used to determine the last population as a function of the others considered as independent variables

$$P_N = 1 - \sum_{i=1}^{N-1} P_i \tag{49}$$

we now derive the geometrical measure for the RPSE described by coordinates $z = (P_1, P_2, \cdots, P_{N-1}, \alpha_1, \alpha_2, \cdots, \alpha_N)$. According to the prescription eq. (10), the metric tensor $g_{ij}(z)$ induced in the surface results to be partitioned in two blocks, the block on the phases which is diagonal

$$g_{\alpha_i \alpha_i} = P_i \text{ for } i \neq N, \qquad g_{\alpha_N \alpha_N} = 1 - \sum_{i=1}^{N-1} P_i \tag{50}$$

and the block on the populations

$$g_{P_i P_j} = \frac{1}{4P_i} \delta_{ij} + \frac{1}{4\left(1 - \sum_{k=1}^{N-1} P_k\right)} \qquad i, j = 1 \div (N-1) \tag{51}$$

In order to calculate the determinant of the metric tensor $g(z) = \det(g_{ij}(z))$, we can employ the following property of determinants. Let us suppose that $A$ is an invertible square matrix and $u$, $v$ are two column vectors; then it can be verified[48] that

$$\det(A + uv^T) = (1 + v^T A^{-1} u) \det(A) \tag{52}$$

By identifying $A$ with the diagonal matrix whose entries are $1/(4P_i)$, $v$ with the vector whose elements are all unitary, and the elements of $u$ with the second term at the r.h.s. of (51), we find the following contribution for the population block

$$\det(g_{P_i P_j}) = \frac{1}{4^{N-1}} \frac{1}{1 - \sum_k^{N-1} P_k} \prod_{k=1}^{N-1} \frac{1}{P_k} \tag{53}$$

while the determinant of the diagonal phase block is directly recovered from eq. (50).

$$\det(g_{\alpha_i \alpha_j}) = \left(1 - \sum_k^{N-1} P_k\right) \prod_{k=1}^{N-1} P_k \tag{54}$$

Therefore the overall determinant of the metric tensor is simply

$$g(z) = \det(g_{P_i P_j}) \det(g_{\alpha_i \alpha_j}) = \frac{1}{4^{N-1}} \tag{55}$$

By taking into account the surface of the hypersphere eq. (4) is $2\pi^N/(N-1)!$, one finally derives the probability density eq. (16) for the RPSE.

The possible states for the FEEE must satisfy the constraints eqs. (13) and (14), which can be employed to determine the last two populations as function of the remaining ones

$$P_N = f_1(P_1, \ldots P_{N-2}) = \frac{E}{E_N - E_{N-1}} - \sum_{i=1}^{N-2} \frac{E_i}{E_N - E_{N-1}} P_i + \frac{E_{k'}}{E_N - E_{N-1}} \left(\sum_{i=1}^{N-2} P_i - 1\right)$$
$$P_{N-1} = f_2(P_1, \ldots P_{N-2}) = 1 - \sum_{i=1}^{N-2} P_i - f_1(P_1, \ldots P_{N-2}) \tag{56}$$

and these equations provide the parametric representation of the FEEE hypersurface. Thus, according to eq. (10), the metric tensor on the populations is given as

$$g_{P_j P_t} = \frac{1}{4P_j} \delta_{jt} + \frac{1}{4f_1} \frac{\partial f_1}{\partial P_j} \frac{\partial f_1}{\partial P_t} + \frac{1}{4f_2} \frac{\partial f_2}{\partial P_j} \frac{\partial f_2}{\partial P_t} \tag{57}$$

Again we can calculate the determinant of this matrix by using a general property of the determinant[48]

$$\det(A + UU^\dagger) = \det(I + U^\dagger A^{-1} U) \det(A) \tag{58}$$

where $A$ is a non-singular square matrix, and $U$ is a rectangular matrix with the same number of rows of $A$. By identifying $A$ with the diagonal matrix with entries $A_{ij} = \frac{1}{4P_j} \delta_{ij}$, $U$ with a $(N-2) \times 2$ matrix with entries $U_{ij} = \frac{1}{\sqrt{4f_j}} \frac{\partial f_j}{\partial P_i}$, we get

$$\det(g_{P_i P_j}) = \left((1 + R_{11})(1 + R_{22}) - R_{12}^2\right) \prod_{j=1}^{N-2} \frac{1}{4P_j} \tag{59}$$

with the following coefficients

$$a_j = \frac{E_{N-1} - E_j}{E_N - E_{N-1}} \tag{60}$$

$$R_{11} = \frac{1}{f_1} \sum_{j=1}^{N-2} P_j a_j^2, \quad R_{22} = \frac{1}{f_2} \sum_{j=1}^{N-2} P_j (1+a_j)^2, \quad R_{12} = \frac{1}{\sqrt{f_2 f_1}} \sum_{j=1}^{N-2} P_j a_j (1+a_j) \tag{61}$$

By employing also the determinant of the phase block eq. (54), finally we derive the FEEE volume element

$$dV = \frac{1}{2^{N-2}} \left[ \sum_{j=1}^{N-2} P_j (1+a_j) a_j - \left( \frac{E - E_{N-1}}{E_N - E_{N-1}} \right)^2 + \left( \frac{E - E_{N-1}}{E_N - E_{N-1}} \right) \right]^{1/2} dP_1 ... dP_{N-2} d\alpha_1 ... d\alpha_N \tag{62}$$

from which the FEEE probability density eq. (17) follows directly.

## APPENDIX B: Minimization of the information functional

The information functional eq. (28) with the distribution function eq. (38) can be written as

$$I[W_{FEEE,II}] = \int d\eta_1 G_a(\eta_1) \ln G_a(\eta_1) + \int d\hat{\eta} \hat{W}(\hat{\eta}) \ln \hat{W}(\hat{\eta}) = I_1(a) + \hat{I}[\hat{W}] \tag{63}$$

where $\hat{W}(\hat{\eta}) = \hat{W}_{FEEE,II}(\hat{\eta})$. The constraints which have to be satisfied are the normalization of the average populations, eq. (27), a fixed value of the average expectation energy, eq. (33), and the normalization of the probability density:

$$\int d\hat{\eta} \hat{W}(\hat{\eta}) = 1 \tag{64}$$

By introducing suitable Lagrange multipliers $\lambda$, $\mu$ and $\kappa$, the functional to be minimized reads

$$F([\hat{W}], a) = I_1(a) + \hat{I}[\hat{W}] - \lambda \left( \sum_k \langle \eta_k \rangle - 1 \right) - \mu \left( \sum_k \langle \eta_k \rangle E_k - E \right) - \kappa \left( \int \hat{W}(\hat{\eta}) d\hat{\eta} - 1 \right) =$$
$$= I_1(a) - a_1 (\lambda + \mu E_1) + \int \hat{W}(\hat{\eta}) d\hat{\eta} \left[ \ln \hat{W}(\hat{\eta}) - \lambda \sum_{k \neq 1} \eta_k - \mu \sum_{k \neq 1} E_k \eta_k - \kappa \right] + \lambda + \mu E + \kappa \tag{65}$$

where $F([\hat{W}], a)$ has to be considered as a functional of $\hat{W}(\hat{\eta})$ and an ordinary function of the parameters $a = (a_1, a_2)$. The minimization thus requires setting to zero the following derivatives

$$\frac{\partial F}{\partial a_m} = \frac{\partial}{\partial a_m} [I_1(a) - a_1(\lambda + \mu E_1)] = 0 \quad \text{for } m = 1, 2 \tag{66}$$

$$\frac{\delta F}{\delta \hat{W}(\hat{\eta})} = -\ln \hat{W}(\hat{\eta}) - 1 - \lambda \sum_{k \neq 1} \eta_k - \mu \sum_{k \neq 1} E_k \eta_k - \kappa = 0 \tag{67}$$

From eq. (67) one obtains

$$\hat{W}(\hat{\eta}) = \exp\left[-(1+\kappa) - \lambda \sum_{k \neq 1} \eta_k - \mu \sum_{k \neq 1} E_k \eta_k\right] \tag{68}$$

which, by taking into account the normalization condition eq. (64), can be written in the normalized form

$$\hat{W}(\hat{\eta}) = \prod_{k \neq 1}^{N} (\lambda + \mu E_k) e^{-(\lambda + \mu E_k)\eta_k} \tag{69}$$

so leading to the following averages

$$\langle \eta_k \rangle = \frac{1}{(\lambda + \mu E_k)} \qquad k \neq 1 \tag{70}$$

Lagrange multipliers $\lambda$ and $\mu$ are calculated by evaluating according to eq. (71) the constraint eq. (27) of population normalization and the constraint eq. (33) of expectation energy

$$a_1 + \sum_{k \neq 1}^{N} \frac{1}{(\lambda + \mu E_k)} = 1, \qquad E_1 a_1 + \sum_{k \neq 1}^{N} \frac{E_k}{(\lambda + \mu E_k)} = E \tag{71}$$

while parameters $a = (a_1, a_2)$ are calculated from the solution of eq.(66). Thus, by using a scale of the energy such that $E_1 = 0$, the equations to be solved for the specification of the parameters $\lambda$, $\mu$, $a_1$ and $a_2$ are

$$\begin{cases} a_1 + \dfrac{1}{\mu} \sum_{k=2}^{N} \dfrac{1}{z + E_k} = 1 \\ \dfrac{(N-1)}{\mu} + \dfrac{z}{\mu} \sum_{k=2}^{N} \dfrac{1}{z + E_k} = E \\ \dfrac{\partial}{\partial a_m}\left[\dfrac{1}{2} - \ln\left(\dfrac{a_2}{\pi}\right) - a_1 z \mu\right] = 0 \quad \text{for} \quad m = 1, 2 \end{cases} \tag{72}$$

where $z = \lambda/\mu$. This system of equations can be solved analytically so deriving the following values for the parameters

$$\lambda = 0 \qquad \mu = \frac{N-1}{E} \qquad a_1 = 1 - \frac{E}{N-1} \sum_{k \neq 1}^{N} \frac{1}{E_k} \qquad 1/a_2 = 0 \tag{73}$$

which lead to the distribution specified by eq. (42) and eq. (40).

**Acknowledgements**

The authors acknowledge the support by Univesità degli Studi di Padova through 60% grants.